\documentclass[prd,10pt,aps,twocolumn,superscriptaddress,floatfix,notitlepage,nofootinbib,amssymb,amsmath]{revtex4-1}
\usepackage{epsfig}

\newcommand{\beqn}{\begin{eqnarray}}
\newcommand{\eeqn}{\end{eqnarray}}
\newcommand{\eq}[1]{(\ref{#1})}

\newcommand{\cO}{{\cal O}}
\newcommand{\cS}{{\cal S}}

\newcommand{\cZ}{{\cal Z}}

\newcommand{\cA}{{\cal A}}
\newcommand{\cD}{{\cal D}}

\newcommand{\lat}{{\mathrm{lat}\,}}
\newcommand{\bx}{\boldsymbol {x}}
\newcommand{\by}{\boldsymbol {y}}
\newcommand{\vx}{\vec {x}}
\newcommand{\vy}{\vec {y}}
\newcommand{\vq}{\vec {q}}
\newcommand{\Cas}{{\mathrm{Cas}\,}}
\newcommand{\Z}{{\mathbb Z}}
\newcommand{\bs}{\boldsymbol}
\newcommand{\plane}{{\cal P}_{\cal S}}
\newcommand{\avr}[1]{{\left\langle #1 \right\rangle}}
\newcommand{\aavr}[1]{\avr{\!\avr{ #1 }}}

\begin{document}

\title{Casimir effect on the lattice: U(1) gauge theory in two spatial dimensions}

\author{M. N. Chernodub}
\affiliation{Laboratoire de Math\'ematiques et Physique Th\'eorique UMR 7350, Universit\'e de Tours, 37200 France}
\affiliation{Soft Matter Physics Laboratory, Far Eastern Federal University, Sukhanova 8, Vladivostok, 690950, Russia}
\affiliation{Department of Physics and Astronomy, University of Gent, Krijgslaan 281, S9, B-9000 Gent, Belgium}
\author{V. A. Goy}
\affiliation{School of Natural Sciences, Far Eastern Federal University, Sukhanova 8, Vladivostok, 690950, Russia}
\affiliation{Soft Matter Physics Laboratory, Far Eastern Federal University, Sukhanova 8, Vladivostok, 690950, Russia}
\author{A. V. Molochkov}
\affiliation{Soft Matter Physics Laboratory, Far Eastern Federal University, Sukhanova 8, Vladivostok, 690950, Russia}

\begin{abstract}
We propose a general numerical method to study the Casimir effect in lattice gauge theories. We illustrate the method by calculating the energy density of zero-point fluctuations around two parallel wires of finite static permittivity in Abelian gauge theory in two spatial dimensions. We discuss various subtle issues related to the lattice formulation of the problem and show how they can successfully be resolved. Finally, we calculate the Casimir potential between the wires of a fixed permittivity, extrapolate our results to the limit of ideally conducting wires and demonstrate excellent agreement with a known theoretical result.
\end{abstract}

\date{September 8, 2016}

\maketitle

\section{Introduction}

The influence of the physical objects on zero-point (vacuum) fluctuations is generally known as the Casimir effect~\cite{ref:Casimir,ref:Bogdag,ref:Milton}. The simplest example of the Casimir effect is a modification of the vacuum energy of electromagnetic field by closely-spaced and perfectly-conducting parallel plates which leads to attraction of the plates to each other. If the objects are made of real materials, then their intrinsic physical properties affect the energy of zero-point fluctuations in accordance with the Lifshitz theory~\cite{Lifshitz:1956zz}. Generally, the Casimir forces depend on permittivity, permeability, conductivity of the material which, in general, are complex functions of the electromagnetic wave frequency. A calculation of the Casimir energy, in general, is quite involved. 

Even in the case of perfect conductors the analytical calculation of the Casimir forces may be a challenge if the objects are not flat. These forces may be either attractive or repulsive depending on geometrical shape of the conductors. There are various numerical tools to compute Casimir interactions~\cite{Johnson:2010ug} including worldline Monte-Carlo methods~\cite{Gies:2006cq,Gies:2003cv}.

In our paper we would like to join investigation of the Casimir physics using Monte-Carlo methods of lattice gauge theories that are usually applied to nonperturbative studies in particle physics. A similar philosophy albeit with different technical implementations has already been successfully used to study the Casimir forces between ideal conductors in Refs.~\cite{ref:Oleg:CS:compact,ref:Oleg:CS:noncompact,ref:Oleg:zero:noncompact}. 

In order to illustrate our method we compute the Casimir interaction between two parallel wires in an Abelian gauge model in two spatial dimensions in thermodynamic equilibrium. We calculate the zero-point energy for the wires described by a finite static permittivity (dielectric constant). We extrapolate our numerical results to the limit of ideal conductors and demonstrating the feasibility of the method by confronting our findings with known analytical results. We notice that our approach can be easily generalized to study the Casimir energy for materials of various shapes (consistent with the lattice discretization) described by spatially-anisotropic and space-dependent static permittivities $\varepsilon(x)$ and permeabilities $\mu(x)$ at zero and finite temperature in theories with various gauge groups.

The structure of the paper is as follows. In Sect.~\ref{sec:boundaries} we review the implementation of the Casimir boundary conditions for ideal conductors and propose its natural counterpart in the lattice gauge theory. We also discuss a simple lattice implementation of materials characterized by static permittivity $\varepsilon$ and permeability $\mu$. Section~\ref{sec:cQED:3D} is devoted to discussion of the Casimir problem in (2+1) dimensional Abelian gauge theory. We discuss geometrical setup, describe relevant observables, derive the effect analytically in the continuum spacetime and rederive it again in the Euclidean lattice spacetime. The latter formula turns out to be crucial for precise comparison -- done in Section~\ref{sec:simulations} -- of the results of our numerical simulations with the theoretical prediction in perfectly conducting limit. In Sect.~\ref{sec:simulations} we calculate zero-point electromagnetic fields around dielectric wires, study Casimir energy, make various scaling checks and discuss subtleties of determination of Casimir energy in continuum based on the lattice results. Finally, we calculate numerically the dependence of the Casimir energy on  static permittivity $\varepsilon$ of the wires. The last section is devoted to our conclusions.

\section{The method} 
\label{sec:boundaries}

\subsection{Casimir boundary conditions in continuum}

\subsubsection{Abelian gauge theory in (3+1) dimensions}

The Casimir effect is best probed by perfectly conducting metallic surfaces of fixed geometrical shapes.  In (3+1) dimensions a perfectly conducting surface $\cal S$ imposes the following (Casimir) boundary conditions on electromagnetic field~$A_\mu$: 
\beqn
B_\perp(x){\biggl|}_{x \in {\cal S}} = 0\,, 
\qquad
{\bs E}_\parallel(x){\biggl|}_{x \in {\cal S}} = 0\,,
\label{eq:Casimir:continuum}
\eeqn
where $B_\perp(x)$ is the magnetic field normal to the surface at the point $x$ and ${\bs E}_\parallel(x)$ is electric field which is tangential to the surface at the point $x$. The ideal metallic surface imposes three conditions on electromagnetic field as it restricts one magnetic component $B_\perp$ and two electric components ${\bs E}_\parallel$.

For a flat static surface perpendicular to the third axis $x_3$, conditions~\eq{eq:Casimir:continuum} take the simple form: 
\beqn
E_1(x) = E_2(x) = B_3(x) = 0\,, \qquad x \in {\cal S}\,,
\label{eq:EEB1}
\eeqn
which can identically be rewritten in terms of the field-strength tensor
\beqn
F_{\mu\nu} = \partial_\mu A_\nu - \partial_\nu A_\mu\,,
\label{eq:F:munu}
\eeqn
as follows:
\beqn
F^{01}(x) = F^{02}(x) = F^{12}(x) = 0\,, \qquad x \in {\cal S}\,.
\label{eq:F01}
\eeqn

In general, the Casimir conditions \eq{eq:Casimir:continuum} can also be rewritten in the Lorentz-invariant form,
\beqn
n_{\nu} {\widetilde F}^{\mu\nu}(x) = 0  \qquad \mbox{(in $D$=3+1)} \,, 
\label{eq:conditions:Lorentz:4d}
\eeqn
where $n_{\mu}$ is a vector tangential to the world-volume\footnote{A two-dimensional surface has a three-dimensional world-volume which include two spatial coordinates and one time coordinate.} of the metallic surface at the point $x$ and the tilde means the duality operation:
\beqn
{\widetilde F}^{\mu\nu} = \frac{1}{2} \epsilon^{\mu\nu\alpha\beta} F_{\alpha\beta} \qquad \mbox{(in $D$=3+1)} \,.
\label{eq:F:dual}
\eeqn
For a flat static  $(x_1,x_2)$ surface the normal vector is directed along the third direction, $n_{\mu} = \delta_{\mu 3}$
and the boundary conditions for the gauge field~\eq{eq:conditions:Lorentz:4d} are indeed reduced to constraints \eq{eq:EEB1} and \eq{eq:F01}.

\subsubsection{Abelian gauge theory in (2+1) dimensions}

In 2+1 dimensions the Casimir effect is formulated as an interaction between metallic wires in two dimensional spatial volume. The world-surface of a wire is a two-dimensional surface as one dimension comes from the spatial dimension of the wire and another dimension comes from the time coordinate. Moreover, in 2+1 dimensions the dual object to the field strength tensor~\eq{eq:F:dual} is a vector:
\beqn
{\widetilde F}^{\mu} = \frac{1}{2} \epsilon^{\mu\alpha\beta} F_{\alpha\beta} \qquad \mbox{(in 2+1)}\,.
\label{eq:F:dual:3d}
\eeqn
Therefore boundary condition~\eq{eq:conditions:Lorentz:4d} reduces in $(2+1)$ dimensions to the following constraint:
\beqn
n_{\mu} {\widetilde {F}}^{\mu}(x) = 0  \qquad \mbox{(in 2+1)}\,.
\label{eq:conditions:Lorentz:3d}
\eeqn

For a straight static wire parallel to the $x_2$ axis, the normal vector $n_\mu$ is directed along the $x_1$ coordinate, so that the Casimir condition~\eq{eq:conditions:Lorentz:3d} is reduced to the single constraint $E_2(x) = 0$ or to
\beqn
F^{02}(x) = 0\,,
\label{eq:F01:3d}
\eeqn
where $x$ belongs to the world surface of the wire. Equation~\eq{eq:F01:3d} implies that the component of the electric field tangential to the wire should vanish at each point of the wire.

\subsection{Casimir boundaries in lattice gauge theory}

\subsubsection{Lattice Abelian gauge theory: Action}

In order to set up a lattice reformulation of the Casimir boundary conditions~\eq{eq:conditions:Lorentz:4d} and \eq{eq:conditions:Lorentz:3d} let us first consider a simplest possible case given by a compact U(1) gauge model (compact electrodynamics) on the lattice. 

The action of the compact electrodynamics is formulated in $D$-dimensional Euclidean spacetime:
\beqn
S[\theta] = \beta \sum_P \left(1 - \cos \theta_P \right)\,,
\label{eq:S}
\eeqn
where $\beta$ is the lattice coupling constant (to be defined below). The sum in Eq.~\eq{eq:S} goes over all elementary plaquettes $P$ of the lattice. Each plaquette $P$ is defined by the position of one of the corners of the plaquette $x = (x_1, \dots, x_D)$ and by the directions of two orthogonal vectors $\mu < \nu$ in the plaquette plane, $P = \{x,\mu\nu\}$ with $\mu,\nu = 1, \dots D$. The plaquette angle $\theta_P$ is constructed from the elementary link angles $\theta_{x,\mu} \in (-\pi,+\pi)$ which belong to the perimeter $\partial P$ of the plaquette $P$:
\beqn
\theta_{P_{x,\mu\nu}} = \theta_{x,\mu} + \theta_{x+\hat\mu,\nu} - \theta_{x+\hat\nu,\mu} - \theta_{x,\nu}\,.
\label{eq:theta:P}
\eeqn

The partition function of the model is as follows:
\beqn
\cZ = \int \cD \theta \, e^{-S[\theta]}\,, 
\label{eq:Z}
\eeqn
where
\beqn
\int \cD \theta \equiv \prod_{l} \int_{-\pi}^\pi d \theta_l\,,
\eeqn
is the integration measure over the lattice gauge field~$\theta_l$.

The action~\eq{eq:S} is invariant under the $2 \pi$ shifts of the plaquette variable,
\beqn
\theta_P \to \theta_P + 2 \pi n\,, \qquad \quad n \in \Z\,.
\label{eq:theta:P:shifts}
\eeqn
Therefore the $2 \pi$-shifted values of the plaquette strength tensor are physically equivalent to each other indicating that the Abelian group is a compact manifold. 

The angular variable $\theta_{x\mu}$ has a sense of a ``latticized''  Abelian gauge field $A_\mu$, $\theta_{x\mu} = a A_{\mu}(x)$, where $a$ is the lattice spacing (i.e., the length of the elementary link of the lattice). In continuum limit $a \to 0$ the plaquette variable~\eq{eq:theta:P} reduces, for finite values of the gauge field $A_\mu$, to the field strength tensor~\eq{eq:F:munu}
\beqn
\theta_{P_{x,\mu\nu}} = a^2 F_{\mu\nu}(x)\,,
\label{eq:theta:P:exp}
\eeqn
where higher $O(a^4)$ corrections are not shown.

Substituting Eq.~\eq{eq:theta:P:exp} into Eq.~\eq{eq:S}, expanding over $a$ and keeping the leading term only, we get the action of the continuum U(1) gauge theory:
\beqn
S[A] = \frac{1}{4 g^2} \int d^D x\,  F_{\mu\nu}^2 (x)\,,
\label{eq:S:continuum}
\eeqn
where we have identified the sum on the lattice with the integral in a continuum limit, $a^D \sum_x \to \int d^D x$, and also imposed the following relation between the lattice spacing $a$, the lattice coupling constant $\beta$ and the continuum electric charge $g$:
\beqn
\beta = \frac{1}{g^2 a^{4-D}}\,.
\label{eq:beta:g:D}
\eeqn
Notice that the above formula indicates that the dimensionality of the continuum coupling $g$ in $D$ spacetime dimensions is $[g] = {\text{mass}}^{4-D}$. 

In addition to regular photon configurations, the action~\eq{eq:S:continuum} describes also singular topological configurations, Abelian monopoles. We are not going to discuss the monopoles in the present paper. To this end we work in a weak-coupling regime where the monopole density is negligibly small. We leave a detailed discussion of the monopole effects on Casimir forces for a sequel article~\cite{ref:in:preparation}.

\subsubsection{Lattice Abelian gauge theory: Boundary conditions}

The Casimir boundary conditions force  certain components of the field strength tensor to vanish at the metallic surfaces~\eq{eq:conditions:Lorentz:4d} and  \eq{eq:conditions:Lorentz:3d} in $D = 3+1$ and $D = 2+1$, respectively. Due to the linear character of these equations, the same conditions hold in the Euclidean spacetime in $D=4$ and $D=3$ equations, correspondingly. 

In the lattice gauge theory the boundary conditions~\eq{eq:conditions:Lorentz:4d} and  \eq{eq:conditions:Lorentz:3d} correspond to the vanishing of the field strength tensor~\eq{eq:theta:P} -- up to the discrete transformations~\eq{eq:theta:P:shifts} -- at a certain set of the plaquettes $P \in \plane$ that are either touching at or belonging to the world-volume of the metallic surfaces. 

For the sake of simplicity, below we consider examples of a static metallic surface in $D=4$ and a straight wire in $D=3$. The generalization to more complicated surfaces is straightforward.

In $D=4$ Euclidean spacetime we consider a straight plate in $(x_1,x_2)$ plane positioned at $x_3 = 0$. The corresponding boundary conditions  are derived from Eq.~\eq{eq:F01}:
\beqn
\cos\theta_{x,14} = \cos\theta_{x,24}  = \cos\theta_{x,12} = 1\,, \quad (\text{in}\ 4D)\,, \quad
\label{eq:F01:latt}
\eeqn
where $x = (x_1, \dots, x_4)$ and the $x_4$ axis is the Euclidean ``time'' direction. Condition~\eq{eq:F01:latt} is valid for all $x_1$, $x_2$, $x_4$ at fixed $x_3 = 0$. In order to derive Eq.~\eq{eq:F01:latt} we noticed that the condition $F_{\mu\nu}(x) = 0$ implies, because of equivalence~\eq{eq:theta:P:shifts}, the validity of the lattice constraint $\theta_{P_{x,\mu\nu}}  = 2 \pi n$ where $n \in \Z$. The latter constraint is equivalent to the requirement $\cos\theta_{P_{x,\mu\nu}} = 1$ at the corresponding plaquettes hence Eq.~\eq{eq:F01:latt}.

The relations from Eq.~\eq{eq:F01:latt} are implemented on the set of plaquettes $\plane$ which includes, at all timeslices 
\begin{itemize}
\item[(i)] all spatial plaquettes $P_{x,12}$ belonging to the plane; 
\item[(ii)] all spatiotemporal plaquettes $P_{x,14}$ and $P_{x,24}$ which are formed by one link in the Euclidean time direction and another link which belongs to the plane. 
\end{itemize}

In $D=3$ Euclidean spacetime we consider a straight wire in $x_2$ direction positioned at $x_1 = 0$. In the case of an ideal metal, the corresponding boundary condition is given by the lattice version of Eq.~\eq{eq:F01:3d}:
\beqn
\cos\theta_{x,23} = 1\,, \quad (\text{in}\ 3D)\,, \quad
\label{eq:F01:latt:3D}
\eeqn
where the $x_3$ axis is now associated with the Euclidean ``time'' direction.

One of the ways to implement the boundary conditions of the type~\eq{eq:F01:latt} and \eq{eq:F01:latt:3D} is to add a set of Lagrange factors that affect the plaquettes in $\plane$. To this end the standard $U(1)$ action~\eq{eq:S} has to be changed as follows:
\beqn
S_{\lambda}[\theta;\plane] = \sum_P \beta_P(\lambda) \cos \theta_P\,,
\label{eq:S:beta}
\eeqn
where 
\beqn
\beta_P(\lambda) = 
\left\{
\begin{array}{ll}
\beta \,, \quad & P \notin \plane\,, \\[2mm]
\beta + \lambda\,, & P \in \plane\,,
\end{array}
\right.
\label{eq:beta:P}
\eeqn
where $\beta$ is given in Eq.~\eq{eq:beta:g:D} for appropriate dimension~$D$.

The Lagrange multiplier $\lambda$ is then sent to infinity in order to enforce the condition~\eq{eq:F01:latt:3D}. Consequently, the partition function~\eq{eq:Z} of the model in the presence of the Casimir plates becomes as follows:
\beqn
\cZ[\plane] & = & \lim_{\lambda \to + \infty} \cZ_\lambda[\plane]\,, 
\label{eq:Z:Casimir:1}
\\
\cZ_\lambda[\plane] & = & \int \cD \theta \, e^{-S_\lambda[\theta;\plane]}\,.
\label{eq:Z:Casimir:2}
\eeqn

A different type of the boundary conditions based on the lattice Chern-Simons action was proposed both in compact~\cite{ref:Oleg:CS:compact} and non-compact~\cite{ref:Oleg:CS:noncompact} U(1) gauge theories. Another approach -- which was used in a non-compact version of the theory  -- is to put a certain set of link variables $\theta_l$ to zero in a fixed gauge~\cite{ref:Oleg:zero:noncompact}. In our paper we work with explicitly gauge-invariant approach keeping the values of the Lagrange multiplier $\lambda$ finite thus allowing us to simulate a system with a finite permittivity $\varepsilon$. We will also take the limit $\lambda \to \infty$ in order to compare our numerical results with the  expression for the ideal metals~ [Eq.~\eq{eq:V:Cas:R} below].

The advantage of our approach is that we use different parameters $\lambda_{ij}$ different for different orientations $P_{ij}$ of the plaquettes in $\plane$, and keep them finite. In $(3+1)$ dimensions the spatial-temporal components of $\lambda$ may be associated with generally anisotropic components of the dielectric permittivity $\varepsilon_i$ via the relation $\lambda_{i4} \sim \beta \varepsilon_i$. The spatial-spatial components may be related to the component of the diamagnetic permeability $\mu_i$ as follows: $\lambda_{ij} \sim \beta \epsilon_{ijk} \mu^{-1}_k$.

Our approach gives us certain freedom in simulations of Casimir systems with electromagnetic properties close to reality. Indeed, instead of enforcing ideal metal conditions~\eq{eq:F01:latt} at infinite $\lambda$, a finite value of $\lambda$ may allow us to simulate finite values of relative permittivity (dielectric constant) $\varepsilon$ and relative permeability (encoded in diamagnetic constant $\mu$) of the material. In order to illustrate our approach, below we study zero-point interactions of thin wires with a finite permittivity $\varepsilon$.

\subsubsection{Non-Abelian gauge theory: Boundary conditions}

For the sake of completeness here we also formulate the Casimir boundary conditions for SU(N) gauge theory. The non-Abelian analogues of the  conditions~\eq{eq:F01:latt} and \eq{eq:F01:latt:3D} are given by a substitution of the cosines $\cos \theta_P$ by 
\beqn
S_P[U] = 1 - \frac{1}{N} {\text{Re}} \, {\text{Tr}} \, U_P\,,
\eeqn
where $U_{P_{x,\mu\nu}} = U_{x,\mu} U_{x+\hat\mu,\nu} U_{x+\hat\nu,\mu}^\dagger U_{x,\nu}^\dagger$ is the SU(N) plaquette matrix constructed from the $U_{x,\mu}$ link fields. The partition function in the presence of the plates is given by Eqs.~\eq{eq:Z:Casimir:1} and \eq{eq:Z:Casimir:2}, where the non-Abelian action is:
\beqn
S_\lambda[U;\plane] = \sum_P \beta_P S_P[U]\,,
\eeqn
and the plaquette-dependent non-Abelian coupling constant $\beta_P$ is given by Eq.~\eq{eq:beta:P}. The SU(N) lattice coupling $\beta$ is related to the continuum non-Abelian charge $g$ similarly to the Abelian case~\eq{eq:beta:g:D}:  $\beta = 2 N/(g^2 a^{4-D})$.

\section{Casimir problem in 3D Abelian gauge theory}
\label{sec:cQED:3D}

\subsection{Geometrical setup}

In order to test our approach we study the compact U(1) lattice gauge theory in $D=3$ Euclidean dimensions. We consider a symmetric $L_s^3$ cubic lattice which corresponds to a zero-temperature theory. We impose the periodic boundary conditions at the opposite sides of the lattice along all three directions. According to Eq.~\eq{eq:beta:g:D} the lattice gauge coupling is given by 
\beqn
\beta = \frac{1}{g^2 a}\,.
\label{eq:beta:g:D3}
\eeqn

\begin{figure}[!thb]
\begin{center}
\vskip 3mm
\includegraphics[scale=0.35,clip=true]{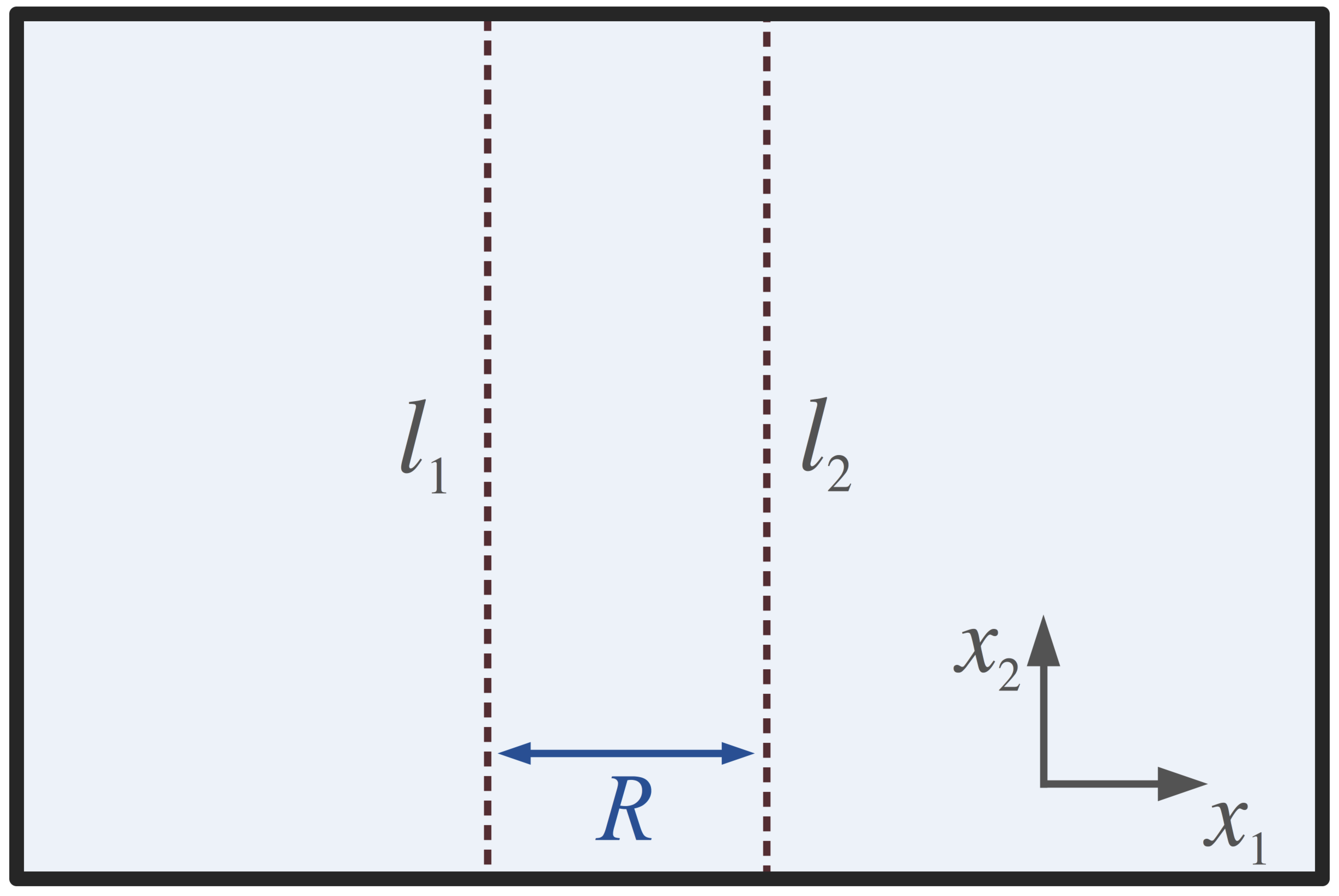}
\end{center}
\vskip -2mm 
\caption{The Casimir problem in two spatial dimensions.}
\label{fig:geometry:plane:2d}
\end{figure}

Our physical setup is shown in Fig.~\ref{fig:geometry:plane:2d}. We consider the simplest case of two parallel static straight wires along the direction $x_2$ separated by the distance $R = |l_2 - l_1|$. The wires are located at positions $x_1 = l_1$ and $x_1 = l_2$. Due to the periodic boundary conditions the wires divide the $x_1$ axis into two intervals, $R$ and $L_s - R$. Therefore, all calculated $R$-dependent quantities (potentials, energy densities, fields, etc) should be invariant under the spatial flip $R \to L_s - R$.

\begin{figure}[!thb]
\begin{center}
\vskip 3mm
\includegraphics[scale=0.35,clip=true]{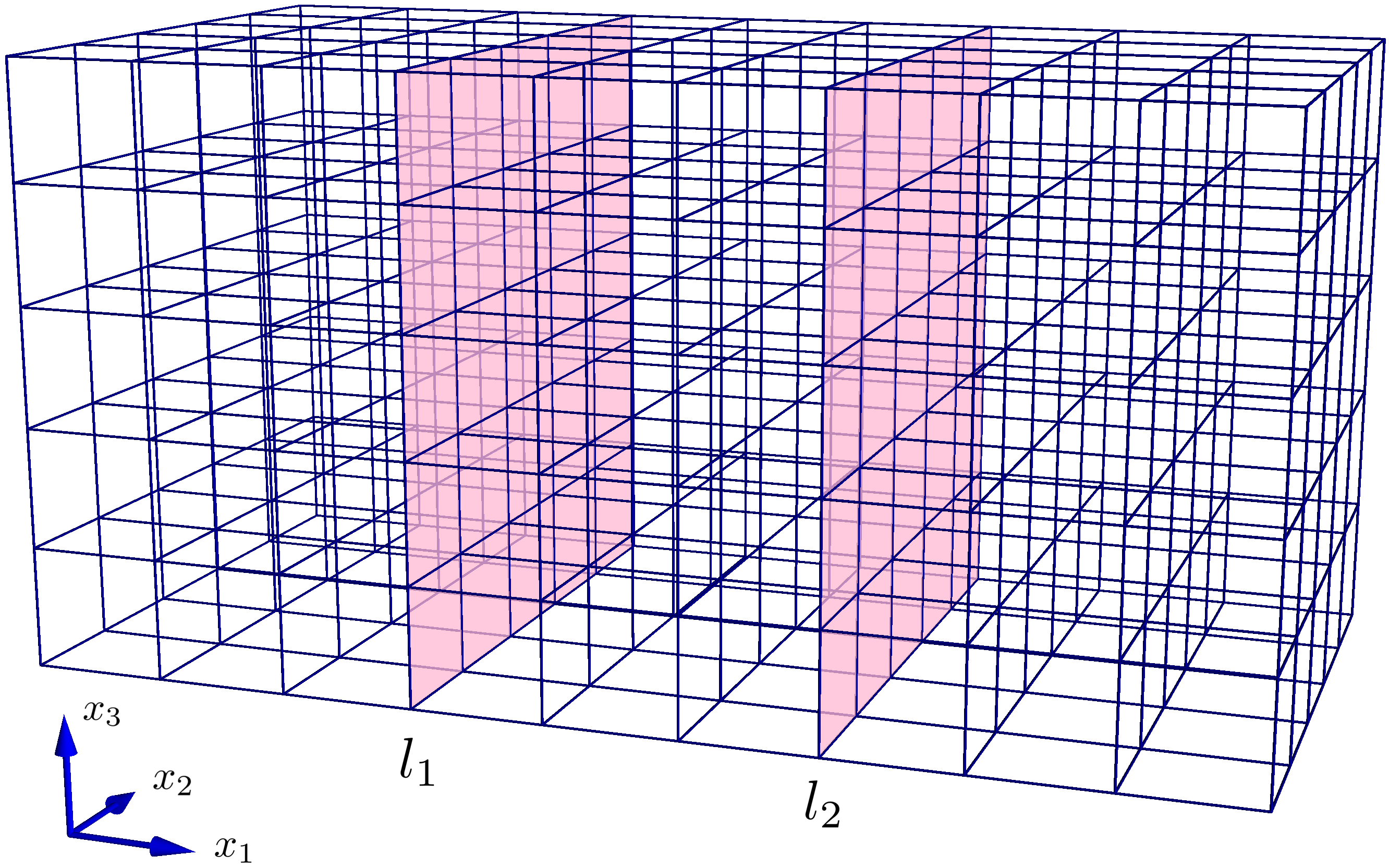}
\end{center}
\vskip -5mm 
\caption{The geometry of the Casimir problem in two spatial dimensions. The shadowed planes indicate the plaquettes $\plane$ where the boundary condition~\eq{eq:F01:latt:3D} is implemented.}
\label{fig:geometry:plane:3d}
\end{figure}

The lattice Casimir condition~\eq{eq:F01:latt:3D} for ideally conducting wires implies vanishing of the ``23''-component of the lattice field strength tensor at the plaquettes which belong to the world-surfaces of the Casimir wires. These plaquettes are visualized in Fig.~\ref{fig:geometry:plane:3d}.  In order to impose the Casimir condition we simulate the compact gauge model with the action~\eq{eq:S:beta}. The coordinate-dependent gauge coupling $\beta_P$, reflects the boundary conditions at the plates~\eq{eq:beta:P}:
\beqn
\beta_{P_{x,\mu\nu}} (\varepsilon) = \beta \bigl[1 + (\varepsilon - 1)\, & & (\delta_{\mu,2} \delta_{\nu,3} - \delta_{\mu,3} \delta_{\nu,2}) \nonumber \\
& & \cdot \left(\delta_{x,l_1} + \delta_{x,l_2}\right)\bigr]\,.
\label{eq:beta:P:3d}
\eeqn
At the plaquettes belonging to the world surfaces of the wires the lattice coupling constant is equal to $\beta_P = \varepsilon \beta $ while outside of the wires, at the bulk majority of the lattice plaquettes, the coupling constant is $\beta_P = \beta$. In terms of our notations of Eq.~\eq{eq:beta:P}, $\lambda = (\varepsilon - 1) \beta$. The quantity $\varepsilon$ is the static permittivity of the material. Below we study the dependence of the Casimir forces on the value of the static permittivity $\varepsilon$. 

In our simulations we realize the case of perfectly conducting wires by taking eventually the limit of large permittivity $\varepsilon \to \infty$. This is possible because in two spatial dimensions the magnetic permeability does not exist and a wire with infinite static dielectric permittivity affects the electromagnetic field in the same way as an ideal metal (cf. Section 5.1 of Ref.~\cite{ref:Bogdag}). Mathematically, in the limit of large dielectric permittivity $\varepsilon \to \infty$  a component of the electric field parallel to the wire vanishes~\eq{eq:F01:3d} thus mimicking an ideal metal.

\subsection{Observables}

The most important observable associated with the Casimir effect is the energy-momentum tensor of the gauge field~\eq{eq:S:continuum},
\beqn
T^{\mu\nu} = - \frac{1}{g^2}F^{\mu\alpha} F^\nu_\alpha + \frac{1}{4 g^2} \eta^{\mu\nu} F_{\alpha\beta} F^{\alpha\beta}.
\eeqn
In Minkowski spacetime the energy density of the gauge field is:
\beqn
T^{00} = \frac{1}{2 g^2} \left(B_z^2 + E_x^2 + E_y^2 \right),
\label{eq:T00:M}
\eeqn
where we took into account the fact that in the spacetime with the metric $(+,-,-)$ one has $F_{01} = E_x$,  $F_{02} = E_y$ and $F_{12} = - B_z$.

From Eq.~\eq{eq:T00:M} we conclude that the Euclidean energy-momentum tensor has the following form:
\beqn
T^{00}_E = \frac{1}{2 g^2} \left(B_z^2 - E_x^2 - E_y^2 \right)\,,
\label{eq:T00:E}
\eeqn
since as we pass from the Minkowski space to the Euclidean space the terms with electric field in Eq.~\eq{eq:T00:M} change their signs, $E_x^2 \to - E_x^2$  and $E_y^2 \to - E_y^2$, while the one with the magnetic field remains intact, $B_z^2 \to B_z^2$. Therefore the 

The rotational symmetry of the problem -- evident from Fig.~\ref{fig:geometry:plane:3d} -- imposes the following constraint on the expectations values:
\beqn
\avr{B_z^2} = \avr{E_x^2}\,.
\label{eq:invariance}
\eeqn
Indeed, the problem is invariant under a $\pm \pi/2$ rotation about the $x \equiv x_1$ axis. The rotation interchanges the axes $x_{2} \leftrightarrow \pm x_3$, and leads to the  transformation of the field strengths: $B_z  \leftrightarrow  \pm E_x$ thus enforcing Eq.~\eq{eq:invariance}.

From Eqs.~\eq{eq:T00:E} and \eq{eq:invariance} we get the normalized energy density:
\beqn
{\cal E}_R(x) & 
= & \avr{T^{00}_E(x)}_{R} - \avr{T^{00}_E(x)}_{0} \nonumber \\
& = & \frac{1}{2 g^2} \left(\avr{E_y^2}_0 - \avr{E_y^2(x)}_R \right)\,,
\label{eq:E:norm}
\eeqn
where the subscript ``0'' indicates that the expectation value is taken in the absence of the wires while the subscript ``$R$'' means that the corresponding average is taken in the presence of the wires separated by the distance $R$. By construction, the energy density \eq{eq:E:norm} is free from ultraviolet divergencies that emerges in the limit $a \to 0$.

The normalized energy density~\eq{eq:E:norm} is a local quantity which is equal to a change in the energy density of vacuum fluctuations that appears due to the presence of the wires. In the geometry of our problem the energy density~\eq{eq:E:norm} depends only on the coordinate $x_1$, which is transverse to the wires themselves. Therefore it is natural to introduce the (Casimir ) energy density per unit length of the wires:
\beqn
V_{\Cas}(R) = \int\limits_{-\infty}^{+\infty} d x_1\, {\cal E}_R(x_1) \equiv - \frac{1}{2 g^2} \aavr{E_y^2}\,,
\label{eq:V:Cas}
\eeqn
which is a finite quantity both in ultraviolet and infrared limits. In Eq.~\eq{eq:V:Cas} the quantity
\beqn
\avr{\!\avr{\cO(x)}\!} = \int d x_1 \left[ \avr{\cO(x)}_R - \avr{\cO}_0 \right]\,,
\label{eq:avr:O}
\eeqn
corresponds to the excess of the expectation value of the operator $\cO$ evaluated per unit length of the wires.

In the lattice regularization 
\beqn
\avr{\!\avr{\cO(x)}\!}_{\lat} =\sum_{x_1 = 0}^{L_s - 1} \left[ \avr{\cO(x_1)}_R - \avr{\cO}_0 \right]\,,
\label{eq:avr:O:lat}
\eeqn
and the lattice Casimir energy density per unit length of the wires~\eq{eq:V:Cas} takes the following compact form:
\beqn
V^{\lat}_{\Cas}(R) = \beta \aavr{\cos \theta_{23}},
\label{eq:V:Cas:lat}
\eeqn
where the lattice coupling $\beta$ is given in Eq.~\eq{eq:beta:g:D3}.

\subsection{Casimir energy: continuum vs. lattice}

\subsubsection{General remarks}

A photon in two spatial dimensions has only one physical degree of freedom. A field, corresponding to this degree of freedom should vanish at the wire. Therefore, it is natural to expect that the Casimir energy in a monopole-free U(1) gauge theory coincides with the one of the free real-valued scalar field with a Dirichlet boundary condition imposed on the field at the wire. The latter energy is known to be as follows~\cite{ref:Wolfram:Ambjorn}
\beqn
V_{\mathrm{Cas}}(R) =
- \frac{\zeta(3)}{16 \pi} \frac{1}{R^2}\,,
\label{eq:V:Cas:R}
\eeqn
where $\zeta(x)$ is the zeta-function with $\zeta(3) \approx 1.20206$. The aim of this section is to derive a lattice version of zero-point (Casimir) energy density between two static straight wires~\eq{eq:V:Cas:R} in the non-compact, monopole-free U(1) gauge theory. To this end we rederive Eq.~\eq{eq:V:Cas:R} in continuum spacetime and then briefly repeat the derivation on the lattice in a weak-coupling regime where the monopole density is negligibly small. The case with monopoles will be treated in details in Ref.~\cite{ref:in:preparation}. 

\subsubsection{Zero-point energy in continuum limit}

\noindent{\sl{Boundary conditions in integral form.}}
A perfect infinitely thin metallic wire forces the electric field along the wire to vanish~\eq{eq:F01:3d} at any point of its world trajectory. In 3 spacetime dimensions the world-trajectory of a wire is a two-dimensional surface $S$ which can be parametrized by a vector ${\bar \bx} = {\bar \bx}(\tau,\xi)$. Here $\tau$ and $\xi$ are time-like and space-like parameters. For example, one can use the following parametrization for a pair of two static wires placed at $x^\pm_1 = \pm R/2$:
\beqn
{\bar \bx}_\pm(\tau,\xi) \equiv (x_1, x_2, x_3) = \left(\pm \frac{R}{2}, \xi,\tau\right)\,,
\label{eq:bx:pm}
\eeqn
where the subscript ``$\pm$'' parameterizes the left/right segments of the wires.

A surface $S$ can be described by the antisymmetric  tensor function
\beqn
s_{\mu\nu}(\bx) = \int d \tau \int d \xi \frac{\partial {\bar x}_{[\mu,}}{\partial \tau} \frac{\partial {\bar x}_{\nu]}}{\partial \xi} 
\delta^{(3)}\Bigl(\bx - {\bar \bx}(\tau,\xi)\Bigr), \quad
\label{eq:s:munu:gen}
\eeqn
where $a_{[\mu,} b_{\nu]} = a_\mu b_\nu - a_\nu b_\mu$. For the parallel static straight wires~\eq{eq:bx:pm} the world surface~\eq{eq:s:munu:gen} has the following form:
\beqn
s^\pm_{\mu\nu}(\bx) = \left(\delta_{\mu,2} \delta_{\nu,3} - \delta_{\nu,3} \delta_{\mu,2} \right) \delta(x_1 \mp R/2)\,.
\label{eq:s:pm}
\eeqn

The Casimir condition~\eq{eq:F01:3d} can conveniently be rewritten in a covariant form with the help of the quantity~\eq{eq:s:munu:gen}:
\beqn
F^{\mu\nu}(\bx) s_{\mu\nu}(\bx) = 0\,.
\label{eq:F:0:cov}
\eeqn
For the static straight wires one automatically gets from Eqs.~\eq{eq:s:pm} and \eq{eq:F:0:cov}:
\beqn
F_{23}(\pm R/2, x_2, x_3) = 0\,.
\label{eq:F23:0}
\eeqn

In the path-integral formalism the Casimir condition~\eq{eq:F:0:cov} can be implemented with the help of a $\delta$ functional which can formally be written as follows:
\beqn
\delta_\cS[F] = \prod_{\bx} \delta\Bigl(F^{\mu\nu}(\bx) s_{\mu\nu}(\bx)\Bigr)\,.
\label{eq:delta:S}
\eeqn
The infinite product of the $\delta$ functions~\eq{eq:delta:S} can be implemented with the help of the functional integration over the Lagrange multiplier $h$:
\beqn
\delta_\cS[F] & = & \int \cD h  \exp\left[ \frac{i}{2} \int d^3 x \, h(\bx) F^{\mu\nu}(\bx) s_{\mu\nu}(\bx) \right],
\label{eq:S:F} \\
& \equiv & \int \cD h  \exp\left[ \frac{i}{2} \int d^3 x \, F^{\mu\nu}(\bx) J_{\mu\nu}(\bx;h) \right]\,,
\eeqn
where we have introduced the source tensor:
\beqn
J_{\mu\nu}(\bx;h) =  h(\bx) s_{\mu\nu}(\bx)\,.
\label{eq:J:munu}
\eeqn
In the case of two parallel Casimir plates~\eq{eq:bx:pm} one gets
\beqn
\delta_\cS[F] = \int \cD h_+ \int \cD h_-  \exp\biggl[ i \int d x_2 \int d x_3 \, \nonumber\\ 
\sum_{a=\pm 1} h_a(x_2,x_3)  F_{23}\left(a \frac{R}{2},x_2,x_3\right) \biggr].
\label{eq:delta:S:plates}
\eeqn
The integration under the exponent is taking place along the two-dimensional world surface, and the integrations over the Lagrange multipliers $h_+$ and $h_-$ enforce the Casimir conditions~\eq{eq:F23:0} at $x_1 = + R/2$ and at $x_1 = - R/2$  plates, respectively.

\vskip 5mm

\noindent{\sl{Zero-point energy in continuum.}} In the presence of the Casimir surface $\cS$ the partition function of the photons can be written as follows
\beqn
Z_\cS  & \equiv & e^{ - \cA \cdot V(R)} = \int \cD A  \, e^{ - S[A]} \, \delta_\cS[F]
=  \int \cD h \, Z_\cS[h], \qquad
\label{eq:Z:S}
\eeqn
where $\cD h = \cD h_+ \cD h_-$ and the photon action $S[A]$ is given in Eq.~\eq{eq:S:continuum} with $D=3$. For the static parallel wires~\eq{eq:Z:ph:lambda:1} Eq.~\eq{eq:Z:S} can be used to define the Casimir energy per unit length of the wires:
\beqn
V(R) = - \frac{1}{\cA} \log Z_\cS\,,
\label{eq:Z:W:R}
\eeqn
where $\cA$ is the worldsheet area of each wire ($\cA = T L$ for a long wire of length $L$ which exists time $T$).

The $h$-dependent partition function~\eq{eq:Z:S} can be rewritten as follows:
\beqn
& & Z_\cS[h] = \int \cD A \, \exp\left[ \int d^3 x \, \left(- \frac{1}{4} F^2_{\mu\nu} + i A_\mu J^\mu\right)\right] \nonumber \\
& & = C \exp\left[ - \frac{1}{2} \int d^3 x \, d^3 y \, J_\mu(\bx; h) D(\bx - \by) J_\mu(\by; h) \right]\!, \qquad 
\label{eq:Z:ph:lambda:1}
\eeqn
where we have performed the Gaussian integration over the photon field $A_\mu$. Notice that the source tensor~\eq{eq:J:munu} enters Eq.~\eq{eq:Z:ph:lambda:1} via the conserved vector 
\beqn
J_\mu(\bx; h) = \partial^\nu J_{\mu\nu}(\bx;h)\,, \qquad \partial^\mu J_\mu(\bx;h) = 0\,.
\label{eq:J:mu}
\eeqn
In Eq.~\eq{eq:Z:ph:lambda:1} the quantity $C$ stands for an inessential constant. From now on we omit this and similar constant factors to simplify our notations.

Substituting Eq.~\eq{eq:Z:ph:lambda:1} into Eq.~\eq{eq:Z:S} one gets
\beqn
Z_{\cS} = \int \cD \Lambda \, e^{ - \frac{1}{2} \int d^2 x \, d^2 y \, \Lambda^T(\vx) {\widehat K}(\vx - \vy) \Lambda(\vy)}\,,
\label{eq:Z:lambda:nomon:1}
\eeqn
where we introduced the two-dimensional vector $\vx = (x_2,x_3)$ on the worldsheet of the wires. We also introduced the vector field:
\beqn
\Lambda(\vx) = 
\left(
\begin{array}{c}
h_+ (\vx) \\
h_- (\vx)
\end{array}
\right)\,,
\eeqn
and the matrix
\beqn
{\widehat K}(\vx) & = & \left(\frac{\partial^2}{\partial x_2^2} + \frac{\partial^2}{\partial x_3^2}\right)
\label{eq:K} \\
& & \left(
\begin{array}{ll}
D(0, x_2, x_3) & D(- R, x_2, x_3) \\
D(+ R, x_2, x_3) & D(0, x_2, x_3) 
\end{array}
\right)\,, \nonumber
\eeqn
with the scalar propagator
\beqn
D({\bs x}) = \int \frac{d^3 k}{(2 \pi)^3} \frac{e^{i {\bs k} \bx}}{k^2} = \frac{1}{4 \pi | \bx |}\,,
\label{eq:D}
\eeqn
which obeys the differential equation
\beqn
- \Delta D(\bx) = \delta(\bx)\,,
\eeqn
where $\Delta \equiv \partial_\mu^2$ is the three-dimensional Laplacian.

The Gaussian integral~\eq{eq:Z:lambda:nomon:1} is given  (up to an inessential multiplicative constant) by the following determinant:
\beqn
Z_{\cS} = {\mathrm{det}}^{-1/2}\, {\widehat K}\,.
\label{eq:Z:lambda:nomon:2}
\eeqn
The Casimir energy per unit wire length can be deduced from Eqs.~\eq{eq:Z:W:R} and \eq{eq:Z:lambda:nomon:2}:
\beqn
\cA \cdot V(R) = - \log {\mathrm{det}}^{-\frac{1}{2}}\, {\widehat K} \equiv \frac{1}{2 } {\mathrm{Tr}} \log {\widehat K}.
\label{eq:A:V:R}
\eeqn

In order to evaluate the potential in Eq.~\eq{eq:A:V:R} we notice first that formally this expression can be written as a sum over all eigenvalues~$\kappa_i$
\beqn
{\mathrm{Tr}} \log {\widehat K} \equiv \sum_i \log \kappa_i\,,
\label{eq:Tr:log:formal}
\eeqn
of the operator ${\widehat K}$:
\beqn
{\widehat K} L = \kappa L\,m
\label{eq:eigen:K}
\eeqn
where $L = L(\vx)$ is an eigenvector of the operator ${\widehat K}$. In the explicit form the eigenvalue equation~\eq{eq:eigen:K} reads as follows:
\beqn
\int d^2 y \, {\widehat K}(\vx - \vy) L(\vy) = \kappa L(\vx)\,.
\label{eq:K:L:xy}
\eeqn

Using the integral representation~\eq{eq:D}, the operator $\widehat K$ in Eq.~\eq{eq:K} can be rewritten as follows 
\beqn
{\widehat K}(\vx) = - \int \frac{d^3 k}{(2\pi)^3} \frac{p_2^2 + p_3^2}{p_1^2 + p_2^2 + p_3^2}
\left(
\begin{array}{ll}
1 & e^{-i p_1 R} \\
e^{i p_1 R} & 1 
\end{array}
\right). \quad
\label{eq:K:p}
\eeqn

Next, we represent the eigenvector $L$ in terms of its Fourier transform,
\beqn
L(\vx) = \int \frac{d^2 q}{(2 \pi)^2} L(\vq) e^{i \vq \vx}\,,
\label{eq:L:x}
\eeqn
substitute Eqs.~\eq{eq:K:p} and \eq{eq:L:x} into Eq.~\eq{eq:K:L:xy}  and integrate over $\vy$. We obtain for Eq.~\eq{eq:K:L:xy}:
\beqn
\int \frac{d^2 q}{(2 \pi)^2} \left[{\widehat Q}(R,{\vec q}) - \kappa \right] L(\vq) e^{i \vq \vx} = 0\,,
\label{eq:eigen:1}
\eeqn
where
\beqn
{\widehat Q}(R,{\vec q}) & = & \int \frac{d p_1}{2\pi} \frac{- {\vec q}^{\,2}}{p_1^2 + {\vq}^{\,2}} 
\left(\begin{array}{ll}
1 & e^{-i p_1 R} \\
e^{i p_1 R} & 1 
\end{array}\right) 
\nonumber \\
& = & - \frac{|\vec q|}{2}
\left(\begin{array}{cc}
1 & e^{- |{\vec q}| R} \\
e^{- |{\vec q}| R} & 1 
\end{array}\right)\,.
\label{eq:hat:Q}
\eeqn

Since Eq.~\eq{eq:eigen:1} should be valid for all vectors $\vq$, we arrive to the following equation for the eigenmodes~$L$:
\beqn
\left[{\widehat Q}(R,{\vec q}) - \kappa \right] L(\vq) = 0\,.
\label{eq:eigen:2}
\eeqn
This equation has the following solutions:
\beqn
\kappa_\pm (\vq) = - \frac{|\vq|}{2} \left(1 \pm e^{- |\vq| R}\right)\,.
\eeqn
The solutions are characterized by the discrete index $\pm$ and continuous parameter $\vq$. The phase space associated with these variables is
\beqn
{\mathrm{Tr}}_{\vq} \equiv \cA \int \frac{d^2 q}{(2\pi)^2} \sum_\pm\,,
\eeqn
where $\cA$ is the area in the transverse $\vx \equiv (x_2,x_3)$ plane. Thus, we get for Eq.~\eq{eq:Tr:log:formal}:
\beqn
{\mathrm{Tr}} \log {\widehat K} & \equiv & \sum_i \log \kappa_i = \cA \int \frac{d^2 q}{(2\pi)^2} \log \det {\widehat Q}(R,{\vec q}) \nonumber \\
& = & \cA \int \frac{d^2 q}{(2\pi)^2} \log \left[\frac{\vq^{\; 2}}{4} \left(1 - e^{- 2 |\vq| R}\right)\right]\,.
\label{eq:Tr:log} 
\eeqn 
Then Eq.~\eq{eq:A:V:R} leads us to the following expression for the potential:
\beqn
V(R) = V_{0} + V_{\rm{Cas}}(R)\,,
\label{eq:V:R:sum}
\eeqn
where
\beqn
V_0 = \frac{1}{2} \int \frac{d^2 q}{(2\pi)^2} \log \frac{\vq^{\; 2}}{4}
\label{eq:A:V:R:0}
\eeqn
is the divergent contribution to the potential $V(R)$. Since this contribution does not depend on the distance $R$ between the plates we can safely omit it. The second term in Eq.~\eq{eq:V:R:sum} is the finite Casimir energy
\beqn
V_{\mathrm{Cas}}(R) = \frac{1}{2} \int \frac{d^2 q}{(2\pi)^2} \log \left(1 - e^{- 2 |\vq| R}\right)\,,
\eeqn
The integral can be evaluated explicitly, the final result for the Casimir energy is shown in Eq.~\eq{eq:V:Cas:R}.

\subsubsection{Zero-point energy on the lattice}

The lattice derivation of the zero point energy follows closely the derivation in the continuum limit. On the lattice of the volume $L_s^3$ the lattice momenta are quantized:
\beqn
p_i = \frac{2\pi n_i}{L_s}\,, \qquad n_i = 0, 1, \dots, L_s -1\,.
\eeqn
The lattice counterpart of Eq.~\eq{eq:hat:Q} is as follows:
\beqn
{\widehat Q}_{\lat}(R;n_2,n_3) & = & - \frac{1}{L_s} \sum\limits_{n_1 = 0}^{L_s - 1} \frac{\sum\limits_{i=2}^3 \left(1 - \cos \frac{2\pi n_i}{L_s}\right) }{\sum\limits_{j=1}^3 \left(1 - \cos \frac{2\pi n_i}{L_s}\right)} \nonumber \\
& & \cdot
\left(\begin{array}{ll}
1 & e^{- \frac{2\pi i R n_1}{L_s}} \\
e^{\frac{2\pi i R n_1}{L_s}} & 1 
\end{array}\right)\,,
\label{eq:hat:Q:lat}
\eeqn
where $R = 0, \dots L_s  - 1$.
The zero-point energy is given by the lattice versions of Eqs.~\eq{eq:A:V:R} and \eq{eq:Tr:log}:
\beqn
V_\lat^{\text{th}}(R) = \frac{1}{2} \sum_{n_2 =0}^{L_s - 1} \sum_{n_3 =0}^{L_s - 1} \log \det {\widehat Q}(R; n_2, n_3)\,.
\label{eq:V:R:lat}
\eeqn
The zero-point energy given by Eqs.~\eq{eq:hat:Q:lat} and \eq{eq:V:R:lat} can easily be computed on the lattice. It contains an inessential constant and the physical $R$-dependent term. 

\begin{figure}[!thb]
\begin{center}
\vskip 3mm
\includegraphics[scale=0.55,clip=true]{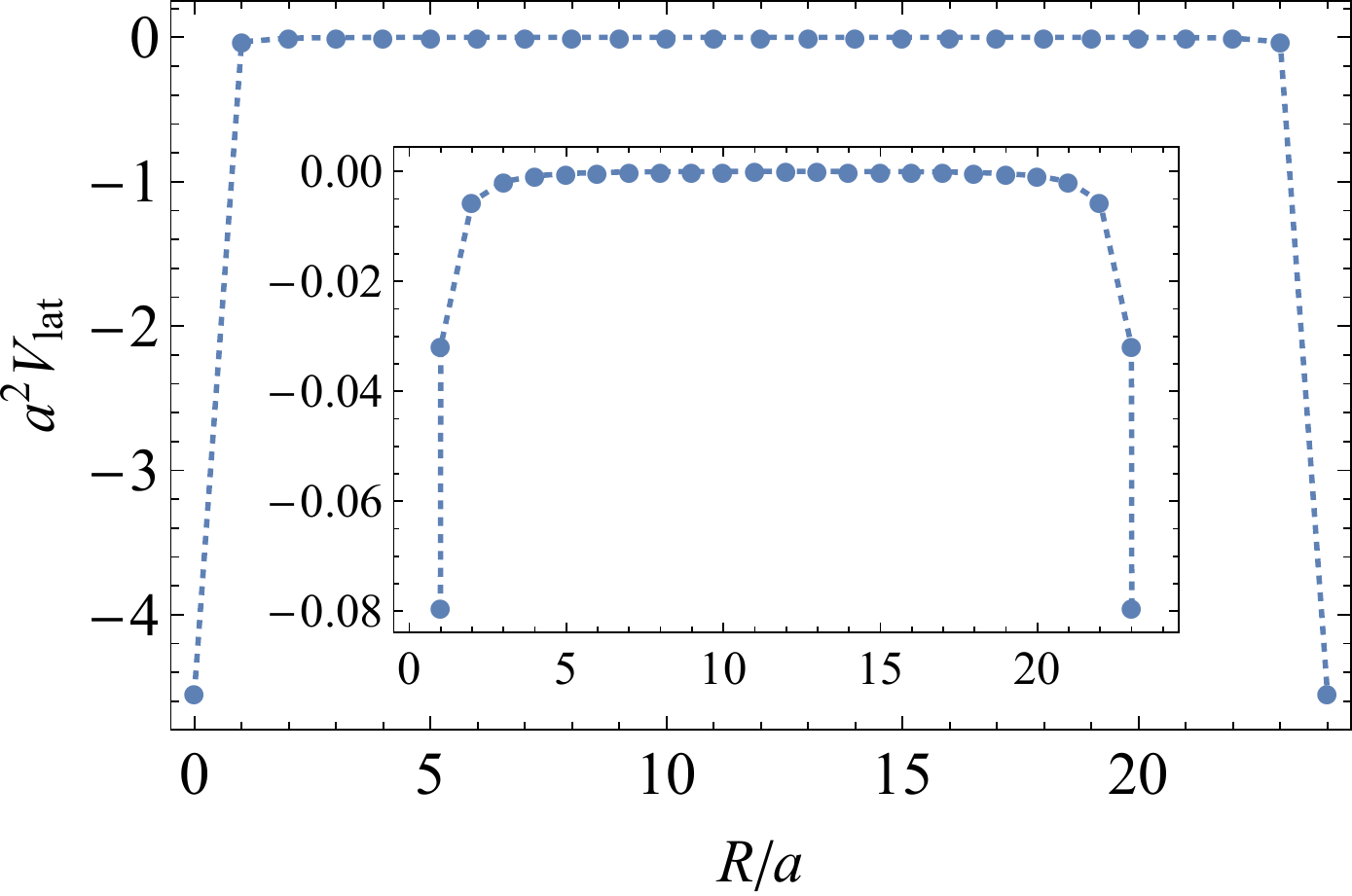}
\end{center}
\vskip -5mm 
\caption{Theoretically calculated zero-point energy~\eq{eq:V:R:lat} between two parallel wires separated by the distance $R$ on $L_s = 24$ symmetric lat\-tice. The inset shows the energy with the $R = 0$ and $R=L_s a$ points excluded.} 
\label{fig:lattice:zero-point}
\end{figure}

As an example, we show the theoretical lattice zero-point energy in Fig.~\ref{fig:lattice:zero-point} for $L_s = 24$ symmetric lattice. Obviously, the energy is invariant under the flips $R \to a L_s - R$. Notice that the depend of the energy on the distance between the plates is very steep at small distances $R$ and it is almost flat at $R \gtrsim 4 a$. Therefore, numerically, the basics Casimir physics can be detected only at small separations between the wires where the ultraviolet lattice artifacts are particularly strong. We will show that this problem is indeed rather serious but it can nevertheless be successfully circumvented.

\section{Casimir energy in simulations}
\label{sec:simulations}

\subsection{Numerical setup}

We first generate configurations with a trivial relative permittivity $\varepsilon = 1$ of the wires, so that at the very fist step the wires are not visible. Then we gradually increase the permittivity of the wires keeping the Wilson gauge coupling $\beta$ fixed. The configurations with higher $\varepsilon$ are generated starting from configurations with lower $\varepsilon$. 

We generate configurations of the gauge fields using a Hybrid Monte Carlo (HMC) algorithm which combines advantages of a molecular dynamics approach and standard Monte-Carlo methods~\cite{ref:Gattringer}. In the molecular dynamics component we use a second-order minimum norm integrator~\cite{ref:Omelyan} with multiple time scales~\cite{ref:Sexton}. The latter allows us to equilibrate the integration errors accumulated at the Casimir planes and outside the planes. This equilibration is particularly important in a limit of high permittivities $\varepsilon \rightarrow \infty$. 
 
We eliminate long autocorrelation lengths in Markov chains of configurations using overrelaxation steps which separate gauge field configurations sufficiently far from each other~\cite{ref:Gattringer}. We also use self-tuning adaptive algorithm in order to control acceptance rate in the HMC in range $[0.70, 0.85]$. The basic parameters of our simulation are presented in Table~\ref{tabl:simparam}.

  \begin{table}[!ht]
    \centering
    \begin{tabular}{|l|c|}
      \hline Trajectories per one value of $\varepsilon$ & $2\ldots 4 \times 10^5$ \\
      \hline Trajectories for thermalization              & $2 \times 10^4$\\
      \hline Overrelaxation steps between trajectories    & $5$\\
      \hline Lattice size & $24^3$ \\
      \hline Range of gauge coupling & $\beta = 3\, \sim\, 7$ \\
      \hline Values of permittivity $\varepsilon$ per single value of $\beta$ & $\approx 20$ \\ 
      \hline
    \end{tabular}
    \caption{Basic simulation parameters.}
    \label{tabl:simparam}
  \end{table}

We use Nvidia graphics processing units (GPU) GTX980 with CUDA architecture as main coprocessors. Since our lattice model involves only nearest-neighbor interactions, the simulations can be parallelized by assigning one GPU thread to perform calculations at each site of the lattice. In order to increase efficiency of the calculations we perform the simulations at the GPUs only thus decreasing data transfer between CPU and GPU.

\subsection{Electromagnetic fields around wires}

In order to characterize the effect of the wires on the local behavior of electromagnetic fields we calculate the lattice quantities (no sum over $ij$ indices is imposed):
\beqn
a^4 \avr{F_{ij}^2(x)}^{\lat}_R = - 2 \left(\avr{\cos\theta_{ij}}_0 - \avr{\cos\theta_{ij}}_R \right)\,.
\label{eq:F2}
\eeqn
The quantity~\eq{eq:F2} is equivalent, up to higher-order $O(a^6)$ terms, to a corresponding component of the electromagnetic field strength squared.

\begin{figure*}[!thb]
\begin{center}
\vskip 3mm
\includegraphics[scale=1,clip=true]{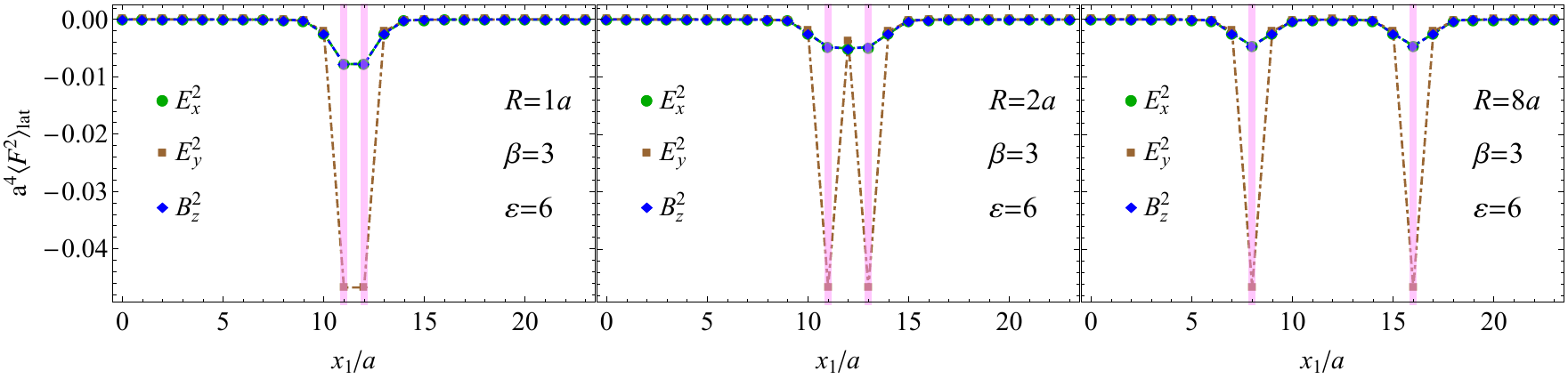}
\end{center}
\vskip -5mm 
\caption{Excess in expectation values of the components of the field strength tensor squared~\eq{eq:F2} vs the coordinate $x_1$ normal to the wires at the lattice coupling $\beta = 3$ and at fixed permittivity~$\varepsilon = 6$. The positions of the wires are marked by the vertical pink lines, the distances between the wires are $R/a = 1,2,8$ for left, middle and right plots, respectively.}
\label{fig:lattice:profiles}
\end{figure*}

In Fig.~\ref{fig:lattice:profiles} we show the components of the field strength~\eq{eq:F2} as the function of the $x \equiv x_1$ coordinate which is normal to the wires. We show the data for the lattice coupling $\beta = 3$ at fixed permittivity~$\varepsilon = 6$ and for three different distances between the wires, $R/a = 1,2,8$. All dimensionful quantities are shown in units of lattice spacing. As expected, each wire strongly suppresses the electric field component parallel to the wire $E_y$.  In accordance with the geometry of the problem, the wires affect also the components $E_x$ and $B_z$ by suppressing their fluctuations $E_x$ and $B_z$ in an equivalent way (the latter is well seen for small separations between the wires). At each wire $\avr{E_x^2}_R= \avr{B_z^2}_R \ll \avr{E_y^2}$ while outside the wires $\avr{E_x^2}_R= \avr{B_z^2}_R \sim \avr{E_y^2}$. The general features of the fields around the wires, shown in Fig.~\ref{fig:lattice:profiles}, are rather universal: the increase of permittivity $\varepsilon$ leads to further suppression of the field fluctuations at the positions of the wires without changing the qualitative shape of the profiles.

Evidently, the larger permittivity $\varepsilon$, the stronger effects of the wires on electromagnetic fields are. In order to characterize the dependence of a quantity $\cO$ on the permittivity~$\varepsilon$ we fitted the corresponding numerical data by the function
\beqn
\cO(\lambda) = \cO_\infty + \frac{C_\cO}{\varepsilon_\cO + \varepsilon}\,,
\label{eq:O:fit}
\eeqn
where $\cO_\infty$, $C_\cO$ and $\varepsilon_\cO>0$ are the fitting parameters.

\begin{figure}[!thb]
\begin{center}
\includegraphics[scale=0.55,clip=true]{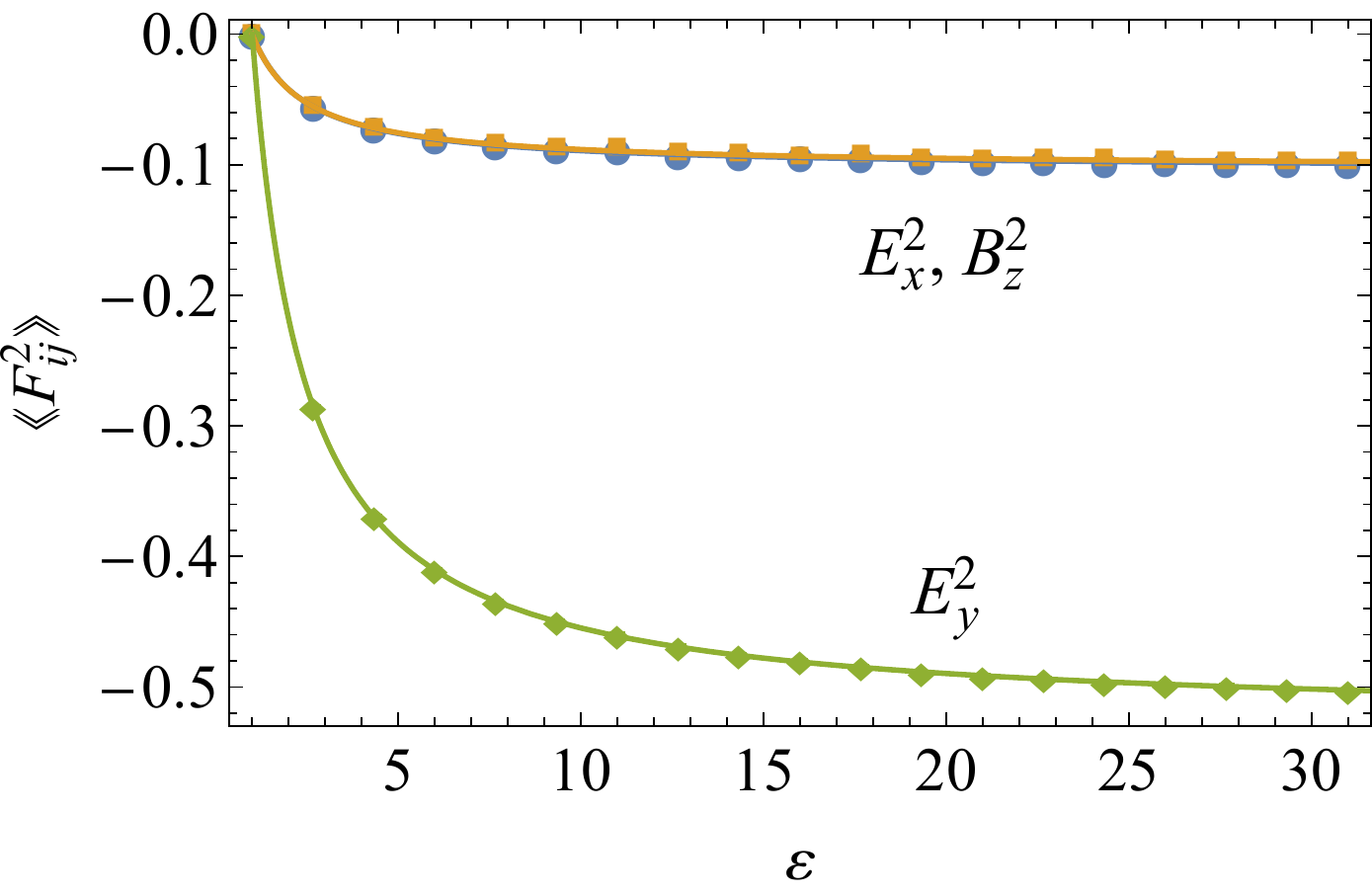}
\end{center}
\vskip -5mm 
\caption{The integral excess in the expectation values of electromagnetic fields~\eq{eq:avr:O} and \eq{eq:avr:O:lat} vs permittivity $\varepsilon$ for $L=24$, $\beta = 3$ and $R=4$. The statistical errors are smaller than the size of the symbols. The lines are the fits by the function~\eq{eq:O:fit}.}
\label{fig:lambda:limit}
\end{figure}

Examples of the normalized mean expectation values of the electromagnetic fields integrated along the coordinate $x_1$ normal to the wires, Eqs.~\eq{eq:avr:O} and \eq{eq:avr:O:lat}, are shown in Fig.~\ref{fig:lambda:limit}. The fits by the function~\eq{eq:O:fit} describe the data very well with $\chi^2/\text{d.o.f.} \approx 1$.

\begin{figure}[!thb]
\begin{center}
\vskip 3mm
\includegraphics[scale=0.55,clip=true]{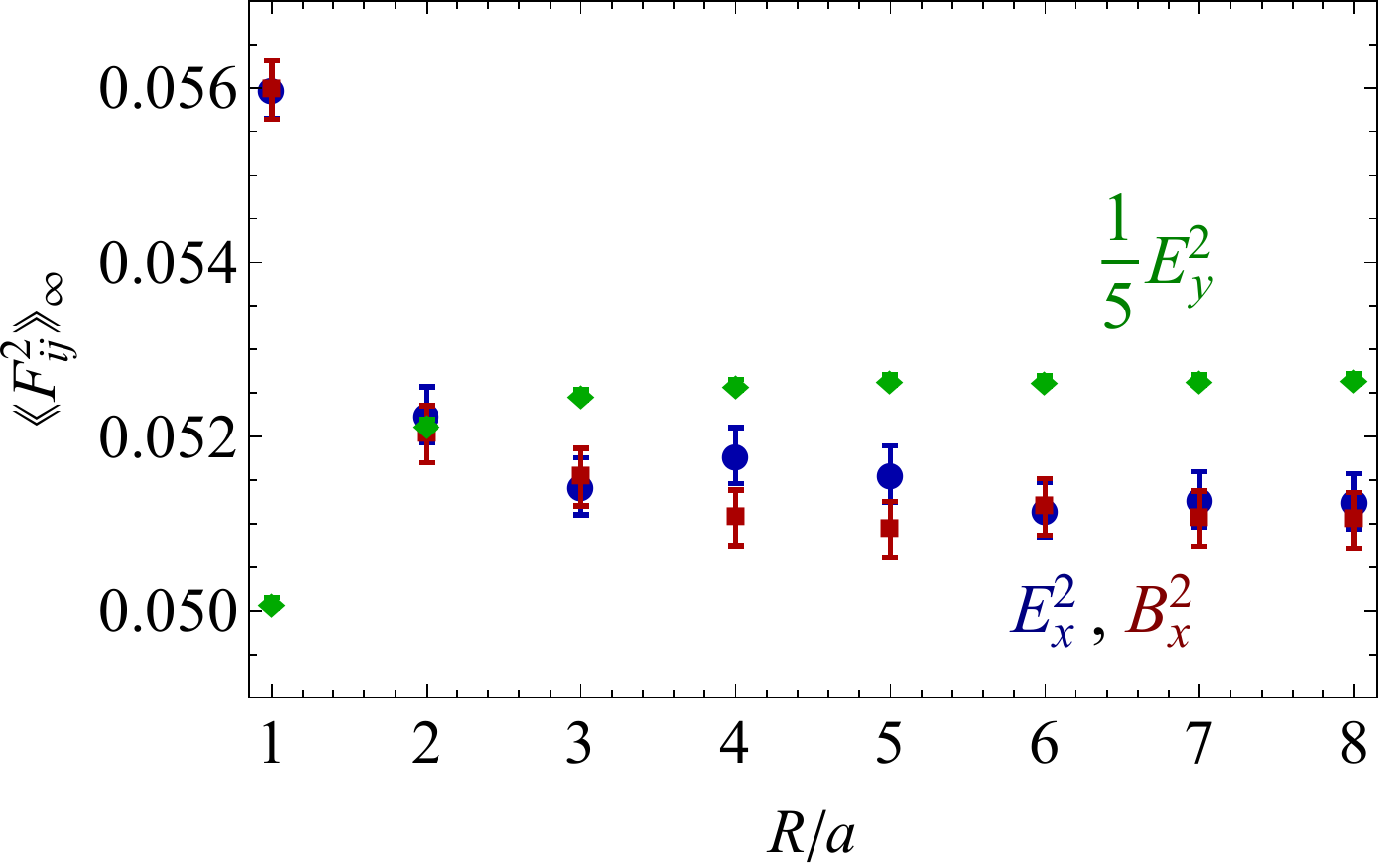}
\end{center}
\vskip -5mm 
\caption{Integral expectation values of electromagnetic fields~\eq{eq:avr:O} and \eq{eq:avr:O:lat} vs the distance between the wires $R$ in the perfect-metal limit  ($\varepsilon \to \infty$) for the lattice $L_s=24$ at $\beta = 3$.}
\label{fig:EB}
\end{figure}

The fits by Eq.~\eq{eq:O:fit} allow us to determine the integral excess in the expectation values of electromagnetic fields in the limit of ideal conductivity. In Fig.~\ref{fig:EB}  we show these quantities as functions of the distance between the wires~$R$. Notice that even in the limit of infinite separation, $R \to \infty$, the integral excesses in electromagnetic fields are nonzero and are independent on the distance $R$ between the wires.

\subsection{Zero-point energy}

\subsubsection{Matching to continuum potential: strong lattice features}

\begin{figure}[!thb]
\begin{center}
\vskip 3mm
\includegraphics[scale=0.55,clip=true]{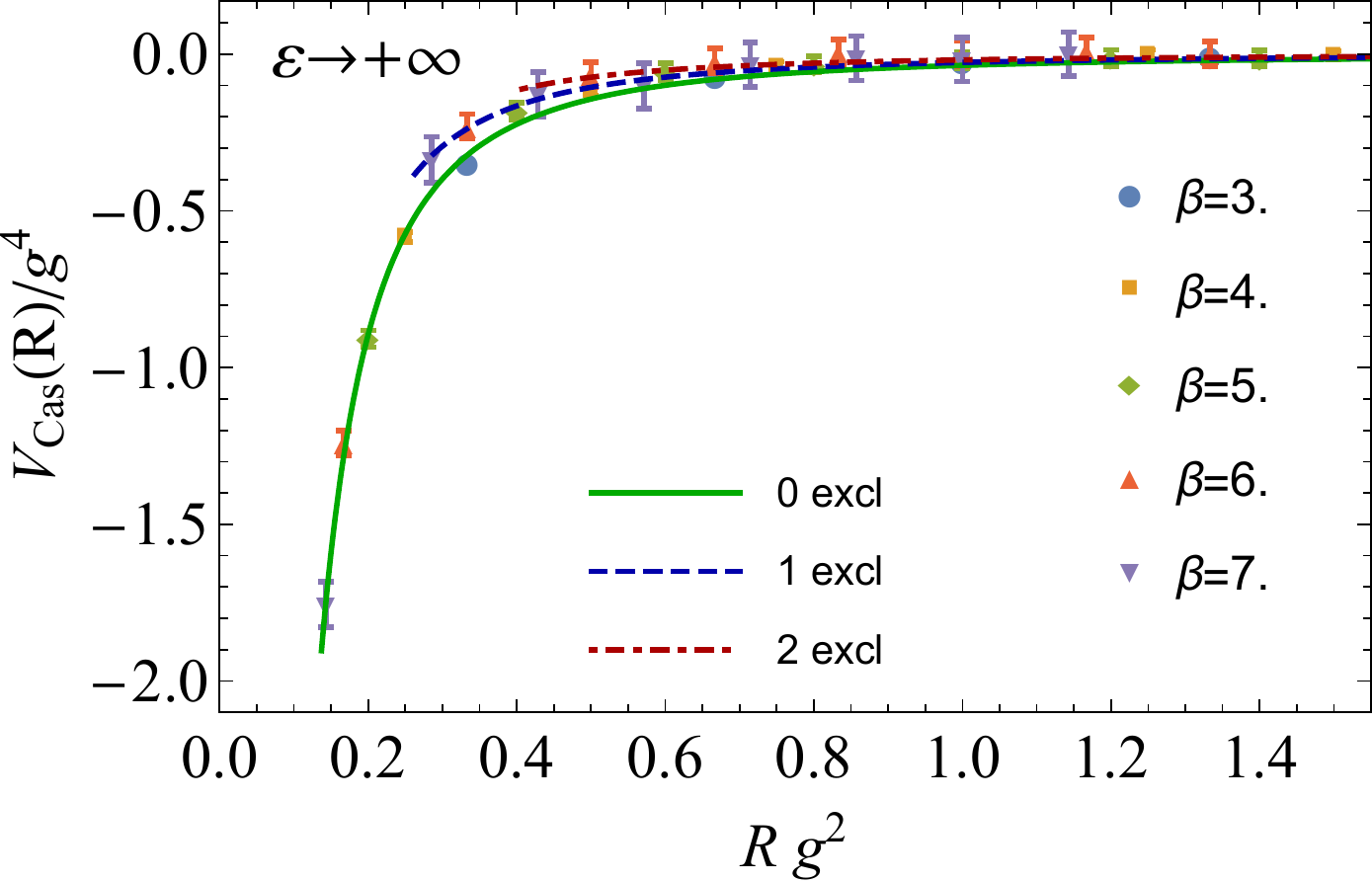}
\end{center}
\vskip -5mm 
\caption{The zero-point potential $V_\Cas$ between perfectly conducting wires separated by the distance $R$ in units of the continuum coupling constant $g$ at various lattice coupling constants $\beta$. The lines represent the fits by Eq.~\eq{eq:V:Cas:R:fit} with zero, one and two first points removed at each $\beta$.}
\label{fig:continuum:fits}
\end{figure}

The integral excess in the $E_y$ component determine the zero-point energy induced by the wires via Eq.~\eq{eq:V:Cas:lat}. The components $E_x$ and $B_z$ do not contribute to the energy density as they exactly cancel each other. In Fig.~\ref{fig:continuum:fits} we show the zero-point potential $V_\Cas$ as a function of the distance $R$, expressed in physical units of the coupling constant $g^2$. We plotted the data for a set of lattice coupling constants $\beta = 3, \dots, 7$, and converted the lattice data from the lattice units (in term of the lattice spacing~$a$) to the physical units (in $g$) using Eq.~\eq{eq:beta:g:D3} for each~$\beta$. Notice that at mentioned values of the lattice coupling $\beta$ the density of the topological lattice configurations, Abelian monopoles, is extremely small,  $\rho_{\mathrm{mon}} \sim 10^{-4} g^6$~\cite{Chernodub:2001ws}. Therefore we neglect the monopoles in our analysis.

The zero-point energy shows a very good scaling in the ultraviolet limit. Indeed, according to Eq.~\eq{eq:beta:g:D3} the physical lattice spacings for $\beta = 3$ and $\beta = 7$ differ more than by a factor of two, and yet the data for $V_\Cas$ expressed in physical units almost coincide with each other, Fig.~\ref{fig:continuum:fits}. 
However, the scaling is not perfect because the data at small -- in lattice units -- wire separations $R/a$ depends very strongly on the distance. Indeed, as we already noticed form Fig.~\ref{fig:lattice:zero-point}, the lattice Casimir potential is very steep at small $R/a$ (where the ultraviolet lattice artifacts are strong) and very flat even at moderate $R/a$ (where the statistical noise is strong). Therefore, we expect that the lattice particularities should be very essential for the Casimir potential, and, consequently, this quantity cannot be  reliably compared to (fitted by) the continuum formula for the zero-point potential~\eq{eq:V:Cas:R}.

In order to check the importance of lattice particularities, we have fitted the lattice zero-point potential $V_\Cas$ by the following continuum function:
\beqn
V_{\mathrm{fit}}(R) = - \frac{C_\lat}{R^2}\,,
\label{eq:V:Cas:R:fit}
\eeqn
where the $C_\lat$ is the sole fitting parameter. In the perfect-metal limit of the theoretical value of $C_\lat$ is~\eq{eq:V:Cas:R}:
\beqn
C_{\mathrm{th}} \equiv C_\infty \equiv \lim_{\varepsilon \to \infty} C(\varepsilon) = \frac{\zeta(3)}{16 \pi} \approx 0.02391 \dots\,.
\label{eq:C:theory}
\eeqn

First, we have fitted by the continuum function~\eq{eq:V:Cas:R:fit} the whole range of the lattice data for $V_\Cas$. Then we have excluded the data at the smallest separation, $R/a$, at each value of $\beta$ and fitted the data again. Next, we removed yet another point $2 R/a$ for each value of $\beta$ and made the fit once again. These three fits are shown in Fig.~\ref{fig:continuum:fits} by the solid, dashed and dot-dashed curves and marked, respectively, by ``0, 1, 2 excl''. For these fits we get different values of the would-be continuum energy $C_\lat = 0.0358(6), \, 0.0264(10), \, 0.0184(14)$ with the quality of the fit defined by  $\chi^2/{\text{d.o.f.}} = 1.45, 0.39, 0.17$, respectively [according to Eq.~\eq{eq:C:theory} the theoretical value for the strength is $C_{\text{th}} \approx 0.0239$]. Thus, we observe that the fit of the lattice data by the theory-inspired continuous function~\eq{eq:V:Cas:R:fit} is strongly dependent on the region of the fit. We conclude that the fitting of the lattice data for the zero-point energy in $(2+1)d$ lattice gauge theory by the continuum function is ambiguous and, therefore, meaningless.

\subsubsection{Matching zero-point energy to the lattice formula}

There are two subtle points which have to be taken into account in order to recover the continuum version of the Casimir potential (the zero-point energy) from the lattice data:
\begin{enumerate} 

\item As we mentioned, we should fit the data for the lattice potential by the ``latticisized'' version~\eq{eq:hat:Q:lat} and \eq{eq:V:R:lat} of the $1/R^2$ potential of continuum theory~\eq{eq:V:Cas:R}.  This procedure allows us to take into account the lattice features of the potential both at short and long distances. The latter takes into account periodicity and mirror ($R \to L_s a - R$) invariance of the potential. 

\item Since our model is compact, the relation between continuum and lattice field strength squared~\eq{eq:F2} is valid up to $O(a^6)$ terms. These corrections are not essential in $\beta \to \infty$ limit where the lattice spacing~\eq{eq:beta:g:D3} is small, $a \ll g^{-2}$. At finite values of the lattice coupling $\beta$ the next-to-the-leading corrections in lattice spacing $a$ can be taken into account by replacing in Eq.~\eq{eq:V:Cas:lat} the Wilson lattice coupling $\beta$ by its Villain counterpart~\cite{Banks:1977cc}:  
\beqn
\beta_V(\beta) = \left[ 2 \log\left( \frac{I_0(\beta)}{I_1(\beta)} \right) \right]\,,
\label{eq:beta:V}
\eeqn
where $I_0$ and $I_1$ are the modified Bessel functions. For reference, the Villain coupling $\beta_V$ is 20\% (7\%) smaller at than the Wilson coupling $\beta$ at $\beta=3$ ($\beta=7$).
\end{enumerate}

We calculate numerically the Casimir potential using Eq.~\eq{eq:V:Cas:lat} in which the Wilson coupling $\beta$ is substituted by its Villain counterpart~\eq{eq:beta:V}, $\beta \to \beta_V$. Then we fit the lattice data for the Casimir potential by the theoretical formula given by Eqs.~\eq{eq:hat:Q:lat} and \eq{eq:V:R:lat}:
\beqn
V_\lat^{\text{fit}}(R) = \frac{C_\lat}{C_{\text{th}}} V_\lat^{\text{th}}(R)\,,
\label{eq:V:lat:fit}
\eeqn
where, for the sake of convenience, we introduced $C_{\text{th}}$, Eq.~\eq{eq:C:theory}. In Eq.~\eq{eq:V:lat:fit} the prefactor $C_\lat$ plays a role of a single fit parameter.

\begin{figure}[!thb]
\begin{center}
\vskip 3mm
\includegraphics[scale=0.55,clip=true]{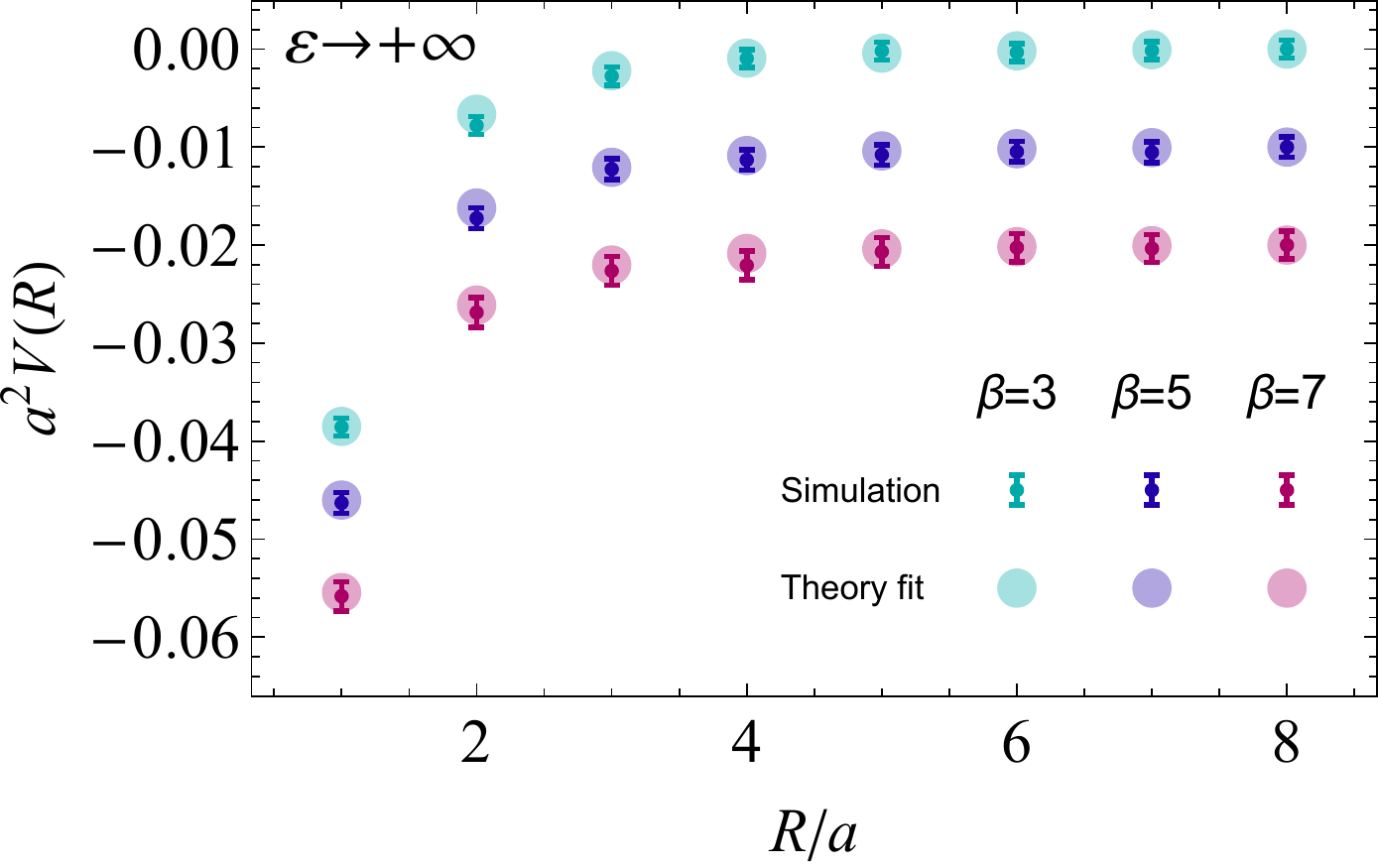}
\end{center}
\vskip -5mm 
\caption{Fits of the numerical data for the zero-point energy by the lattice potential~\eq{eq:V:lat:fit} on the lattice $L_s = 24$ at various~$\beta$. The wires are perfectly conducting.}
\label{fig:lattice:fits}
\end{figure}

The examples of the lattice fits in the ideal-metal limit $\varepsilon \to \infty$ are shown in Fig.~\ref{fig:lattice:fits}. The lattice function~\eq{eq:V:lat:fit} describes the numerical data almost perfectly with $\chi^2/\text{d.o.f} \lesssim 1$. The results are essentially robust against removal one or two points in the ultraviolet region as the corresponding fitting parameters $C_\lat$ coincide with each other within error bars. The latter property highlights the correctness of the chosen fit method. 

\begin{figure}[!thb]
\begin{center}
\includegraphics[scale=0.55,clip=true]{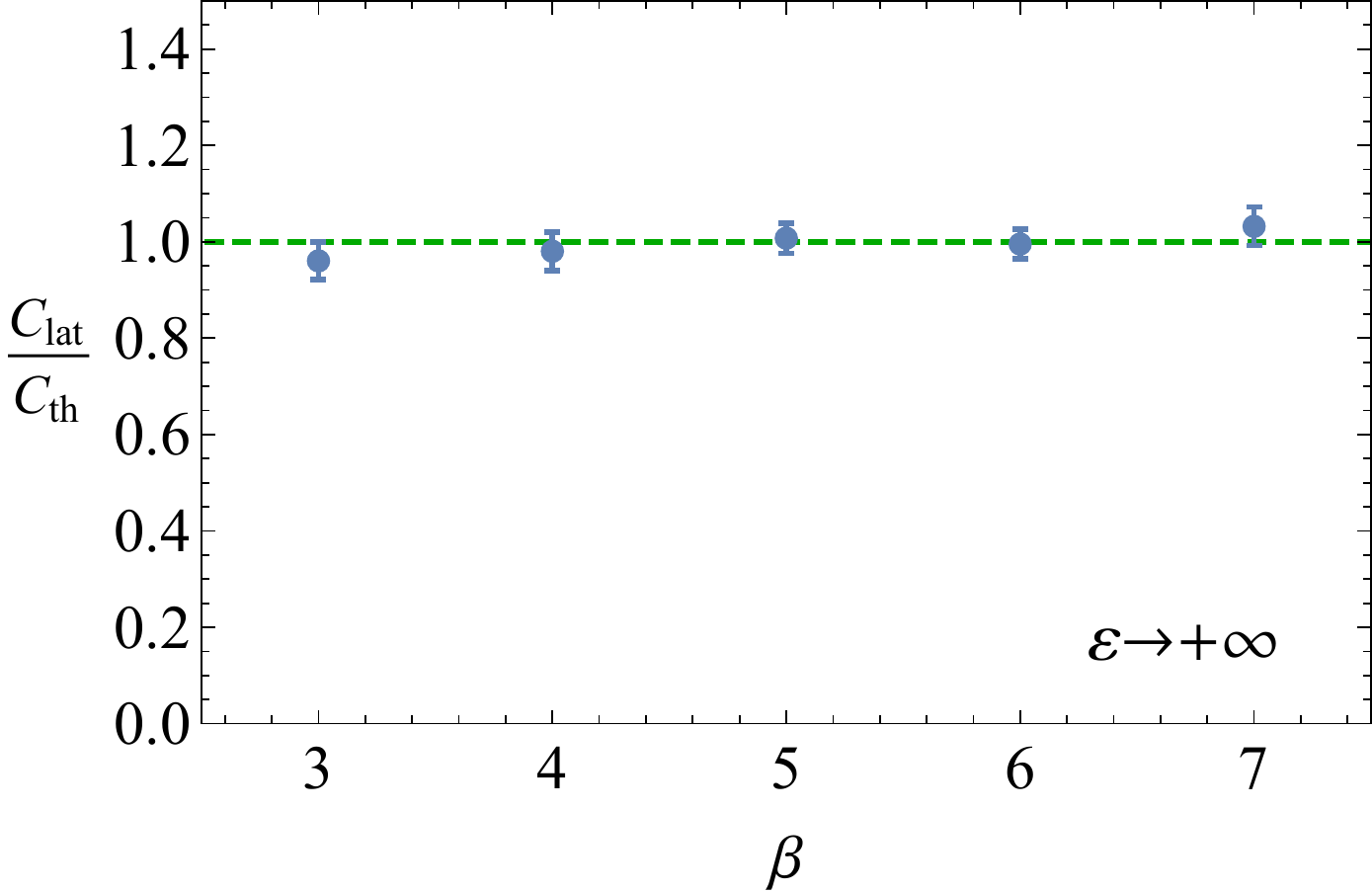}
\end{center}
\vskip -5mm 
\caption{The ratio of numerically simulated vs. theoretically calculated prefactors of the Casimir energy at various lattice coupling constants $\beta$ for perfectly conducting wires. }
\label{fig:accuracy}
\end{figure}

The dependence of the best fit parameter $C_\lat$ on the value of the lattice coupling $\beta$ in the ideal-metal limit is shown in Fig.~\ref{fig:accuracy}. First of all, we notice that the numerical result matches perfectly the theoretical prediction because $C_\lat = C_{\mathrm{th}}$ within small error bars for all studied values of the lattice coupling $\beta$. We may attribute this property to inevitable finite-size corrections which may appear due to Wilsonian cos-type of the chosen action. We also notice from Fig.~\ref{fig:accuracy} that the Casimir effect is independent on the lattice coupling which implies a very good scaling towards the continuum limit. The latter matches well with the theoretical fact that the Casimir energy between ideally conducting plates should not depend on the coupling constant.

\subsubsection{Casimir effect at finite permittivity}

Our approach allows us to study dependence of the Casimir force on the permittivity $\varepsilon$ of the wires. Since in our setup the wires are infinitely thin, the system can be characterized, in a continuum limit, by the following space-dependent permittivity:
\beqn
\varepsilon(x) = 1 + \varepsilon \bigl[ \delta(x - l_1)  + \delta(x - l_2) \bigr]
\label{eq:varepsilon:continuum}
\eeqn
where  $x\equiv x_1$ and $|l_1 - l_2| = R$ according to Fig.~\ref{fig:geometry:plane:2d}.

We have repeated the analysis of the previous section for a wide range of $\varepsilon$. We have found that at finite $\varepsilon$ the data match very well the lattice version~\eq{eq:V:R:lat} of the $1/R^2$ potential. The corresponding continuum counterpart of the zero-point potential is as follows:
\beqn
V_\Cas(R,\varepsilon) = - \frac{C(\varepsilon)}{R^2}\,.
\label{eq:V:Cas:epsilon}
\eeqn

\begin{figure}[!thb]
\begin{center}
\includegraphics[scale=0.55,clip=true]{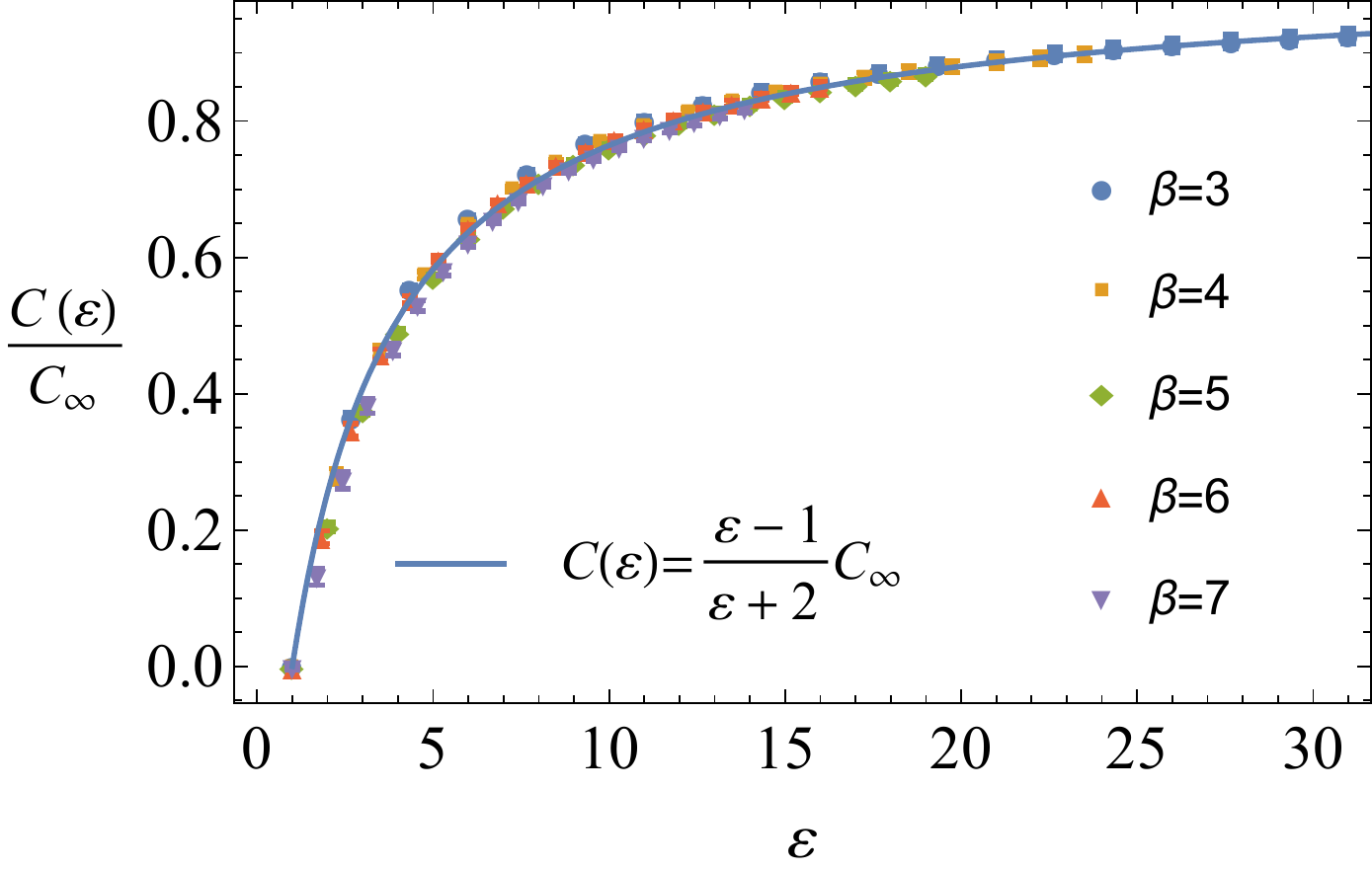}
\end{center}
\vskip -5mm 
\caption{The strength factor $C$ of the zero-point energy~\eq{eq:V:Cas:epsilon} as the function of permittivity $\varepsilon$ of the wires. The solid line corresponds to the function~\eq{eq:C:epsilon}.}
\label{fig:limits}
\end{figure}

In Fig.~\ref{fig:limits} we show the dependence of the strength $C(\varepsilon)$ of the Casimir interaction~\eq{eq:V:Cas:epsilon} on the permittivity of the wires as compared to the strength of the effect in the ideal-metal limit at infinite permittivity. The numerical data  show very good scaling as the data points corresponding to different values of the lattice coupling $\beta$ lie on the same curve. The dependence of the strength factor $C(\varepsilon)$ on permittivity can be very well described by the following function\footnote{A fit of the lattice data by the function $\frac{C(\varepsilon)}{C_\infty} = \frac{\varepsilon + \varepsilon_1}{\varepsilon + \varepsilon_2}$ gives the following best fit parameters: $\varepsilon_1 = - 1.01(1)$ and $\varepsilon_1 = 1.99(7)$.}: 
\beqn
C(\varepsilon) = \frac{\varepsilon - 1}{\varepsilon + 2}  C_\infty\,,
\label{eq:C:epsilon}
\eeqn 
where the factor $C_\infty$ for an ideal conductor is given in Eq.~\eq{eq:C:theory}. The function in Eq.~\eq{eq:C:epsilon}, shown in Fig.~\ref{fig:limits} by the solid line, describes the numerical data almost perfectly. 

\section{Conclusions}

We proposed a simple method to calculate the zero-point (Casimir) vacuum energy using Monte-Carlo methods in the framework of lattice gauge theory. 

The Casimir energy is associated with modification of the zero-point vacuum fluctuations in the presence of physical objects which impose certain boundary conditions on (electromagnetic) fields and/or affect the fields via relative permittivity $\varepsilon$ and/or permeability~$\mu$ of the material. In our numerical method the materials are described by  space- and orientation-dependent gauge coupling~\eq{eq:beta:P} in the lattice action~\eq{eq:S:beta}. 

In order to illustrate our approach we calculated the zero-point energy between two parallel thin wires characterized by a finite static permittivity $\varepsilon$. We carried out our simulations in a weak-coupling regime of an Abelian gauge theory in two spatial dimensions. The appropriate modification of the lattice gauge coupling is given by Eq.~\eq{eq:beta:P:3d}. In the continuum limit the wires are described by the spatially dependent permittivity~\eq{eq:varepsilon:continuum}.

In the limit of an ideally conducting wire, $\varepsilon \to \infty$, our result for the zero-point energy agrees very well with the analytical formula~\eq{eq:V:Cas:R}, Fig.~\ref{fig:accuracy}. At finite values of the relative permittivity of the wires, $\varepsilon \geqslant 1 $ the vacuum energy has the Casimir form~\eq{eq:V:Cas:epsilon} with a modified prefactor~\eq{eq:C:epsilon} shown in Fig.~\ref{fig:limits}. Generally, our lattice data exhibit very accurate ultraviolet scaling. 

Our method is suitable for calculations of the zero-point energy in thermodynamic equilibrium for materials described by (generally, space-dependent) static permittivity $\varepsilon(x)$ and permeability $\mu(x)$. It can be generalized to calculate the vacuum energy for spatially anisotropic materials of various shapes (with an appropriate discretization), at zero and finite temperatures, and for any, including non-Abelian, gauge groups.

\acknowledgments

The work was supported by the Federal Target Programme for Research and Development in Priority Areas of Development of the Russian Scientific and Technological Complex for 2014-2020 (Contract 14.584.21.0017).

\end{document}